\documentclass[sigconf, balance=false]{acmart}

\usepackage{enumitem}
\usepackage[most]{tcolorbox}
\newtcolorbox{promptbox}[1]{
  enhanced, breakable,
  colback=black!3, colframe=black!45,
  boxrule=0.5pt, arc=2pt,
  left=4pt, right=4pt, top=3pt, bottom=3pt,
  fonttitle=\bfseries\small, title=#1
}

\AtBeginDocument{%
  }

\usepackage{mathtools}
\usepackage{bm}
\usepackage{siunitx}
\usepackage{multirow}
\usepackage{makecell}
\usepackage{tabularx}
\usepackage{threeparttable}

\usepackage{algorithm}
\usepackage{algpseudocode}
\usepackage{booktabs}
\newcommand{\ph}[1]{#1}
\usepackage{tabularx}
\usepackage[table]{xcolor}
\definecolor{gainred}{RGB}{180,50,50}

\usepackage{subcaption}

\usepackage{pgfplots}
\pgfplotsset{compat=1.18}
\definecolor{evalblue}{HTML}{2F5D9E}
\definecolor{evalorange}{HTML}{D9822B}
\definecolor{evalgreen}{HTML}{4C9A6A}
\definecolor{evalbarA}{HTML}{B8C4D6}
\definecolor{evalbarB}{HTML}{6F8FBF}
\definecolor{evalshade}{HTML}{F3F5F9}
\colorlet{evalgrid}{black!10}
\begin{document}


\title{EvoRec: A Unified Self-Evolving Agentic Recommender with Dynamic Reflection Trees}

\titlenote{Research team: Kuaishou Local Services Ads Recommendation Foundation Model Team. For inquiries or collaboration, please contact \href{mailto:jianghao11@kuaishou.com}{jianghao11@kuaishou.com}.}

%
%
%


\author{Mengting Li}
\authornote{This work was conducted during Mengting Li's internship at Kuaishou.}
\authornote{Mengting Li and Long Zhang contributed equally to this work.}
\affiliation{%
  \institution{Kuaishou}
  \city{Beijing}
  \country{China}
}
\email{\detokenize{li_mengting@buaa.edu.cn}}

\author{Long Zhang}
\authornotemark[3]
\affiliation{%
  \institution{Kuaishou}
  \city{Beijing}
  \country{China}
}
\email{dragonzhang@mail.ustc.edu.cn}

\author{Hao Jiang}
\authornote{Corresponding author.}
\affiliation{%
  \institution{Kuaishou}
  \city{Beijing}
  \country{China}
}
\email{jianghao11@kuaishou.com}

\author{Sheng Yu}
\affiliation{%
  \institution{Kuaishou}
  \city{Beijing}
  \country{China}
}
\email{yusheng03@kuaishou.com}

\author{Fei Pan}
\affiliation{%
  \institution{Kuaishou}
  \city{Beijing}
  \country{China}
}
\email{panfei05@kuaishou.com}

\author{Peng Jiang}
\affiliation{%
  \institution{Kuaishou}
  \city{Beijing}
  \country{China}
}
\email{jiangpeng@kuaishou.com}

\renewcommand{\shortauthors}{Li et al.}


\begin{abstract}

Conventional recommender systems follow cascaded retrieval, pre-ranking, and ranking pipelines with predefined models and offline-optimized configurations, limiting their ability to adapt to evolving user needs and observed outcomes. Recent agentic recommenders introduce reasoning and tool use but largely focus on isolated stages, lacking both end-to-end decision control and systematic reflection on recommendation failures.
To address these limitations, we propose \textbf{EvoRec}, a unified agentic recommendation framework with dynamic reflection that integrates end-to-end autonomous recommendation with reflection-driven self-evolution. During recommendation, a \textbf{Need Agent} infers structured needs from long-term, short-term, and working memory, a \textbf{Planning Agent} dynamically allocates retrieval budgets between Behavior Continuation and Need Satisfaction while selecting retrieval tools and ranking-fusion weights, and an \textbf{Acting Executor} deterministically executes the resulting plan and records the complete decision and execution trace. After observing user feedback, a \textbf{Dynamic Reflection Tree} localizes failures to need inference, recommendation planning, or prior repair strategies, generates personalized repairs, and verifies their effectiveness through replay. Verified repairs are incorporated into working memory to guide subsequent decisions and further abstracted into reusable strategies shared across users. 
EvoRec thereby establishes a closed loop of Need Understanding $\rightarrow$ Planning $\rightarrow$ Execution $\rightarrow$ Feedback $\rightarrow$ Reflection $\rightarrow$ Repair, enabling continuous self-evolution without updating agent parameters.
Experiments on a Kuaishou local-services advertising dataset show that EvoRec improves Hit@50 by 15.7\% over the strongest baseline. Controlled ablations validate both adaptive agentic decision-making and hierarchical reflection, with the latter improving Hit@50 by 13.1\% for users with prior repair experience. A seven-day online A/B test shows that nearline-deployed EvoRec increases advertising revenue by 4.6\% and CTCVR by 3.5\%, with a 14.5\% revenue gain among high-intent users. These results establish EvoRec as a practical, industrial-scale, unified agentic recommender capable of reflection-driven self-evolution. The EvoRec project is publicly available at \url{https://mtli-buaa.github.io/evorec-project-page/}.
\end{abstract}

%
%
%
%
%
%

\setcopyright{none}
\copyrightyear{2027}
\acmYear{2027}
\acmDOI{}
\acmISBN{}
\settopmatter{printacmref=false}
\renewcommand{\footnotetextcopyrightpermission}[1]{}
\citestyle{acmnumeric}




\maketitle


\begin{figure}[!t]
    \centering
    \includegraphics[width=\linewidth]{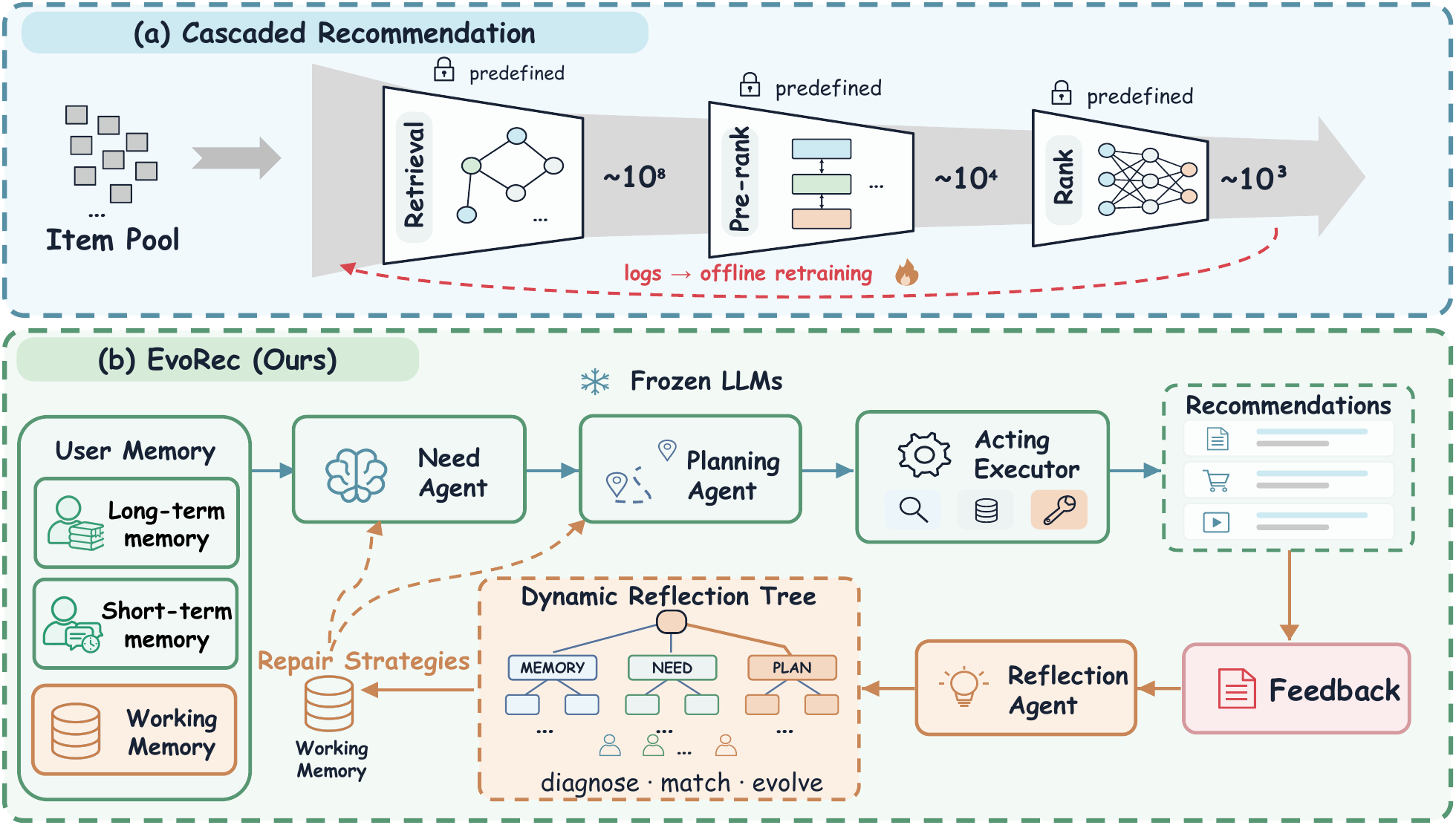}
    \caption{\textbf{Conventional cascaded recommendation vs.\ EvoRec.}
    (a) Conventional recommenders rely on predefined pipelines and offline parameter updates.
    (b) EvoRec integrates autonomous recommendation with reflection-driven self-evolution, using a Dynamic Reflection Tree to diagnose errors and provide personalized repair strategies without updating agent parameters.
    }
    \label{fig:motivation}
\end{figure}

\section{Introduction}
\label{sec:introduction}

Recommender systems play a central role in modern Web platforms, connecting users with relevant content, products, and services.
Traditional recommenders typically adopt multi-stage architectures, in which predefined retrieval strategies generate candidates and ranking models predict user preferences~\cite{kang2018self,yi2019sampling,covington2016deep,jiang2026rest}.
Despite their success, these systems largely rely on predesigned workflows, limiting their ability to autonomously adapt recommendation decisions to evolving user states and observed outcomes.
Recent advances in large language models (LLMs) open new opportunities for recommendation.
Representative studies such as OneReason~\cite{team2026onereason} and OneRec-Think~\cite{liu2025onerec} incorporate reasoning into recommendation models, enhancing preference understanding and item prediction.
However, stronger reasoning within a model does not by itself give the system autonomous control over the entire recommendation process.
Leveraging the semantic understanding, reasoning, and tool-use capabilities of LLMs to move recommendation from passive preference prediction toward autonomous decision-making~\cite{yao2022react} has become a promising research direction.

Following this direction, recent studies have explored agentic recommendation.
InteRecAgent~\cite{huang2025recommender} and ToolRec~\cite{zhao2024let} enable agents to invoke conventional recommendation models as tools, while MACRec~\cite{wang2024macrec} coordinates multiple specialized agents for recommendation.
ChainRec~\cite{li2026chainrec} and RecThinker~\cite{zhang2026recthinker} further explore dynamic tool orchestration and active information acquisition.
Industrial systems such as RecGPT~\cite{yi2025recgpt}, DREAM~\cite{zhang2026dream}, and AgentX~\cite{lao2026agentx} also bring agents into production recommendation pipelines.

However, two limitations remain.
First, these studies mainly address different aspects, ranging from reasoning and tool coordination to pipeline control and algorithm automation, rather than letting agents make actual recommendation decisions end to end, which requires inferring latent needs from dynamic user states, formulating recommendation plans, and coordinating heterogeneous tools accordingly.
Second, recommendation feedback is typically used to update model parameters or as interaction memory~\cite{li2026agenticrec,yi2025recgpt}, without identifying which decision caused an undesirable outcome.
Since a missed conversion may stem from an unrecognized need, an inappropriate plan, or a previously learned strategy, and similar error mechanisms recur across users while effective repairs depend on individual users, such feedback is difficult to turn into reusable experience for subsequent recommendation decisions.

To address these limitations, we propose \textbf{EvoRec}, a unified agentic recommendation framework with hierarchical reflection.
EvoRec formulates recommendation as a decision-making process comprising need understanding, autonomous planning, and plan execution.
Specifically, the \textbf{Need Agent} infers structured user needs from long- and short-term user memory; the \textbf{Planning Agent} allocates the retrieval budget between two complementary paths, \textit{Behavior Continuation} and \textit{Need Satisfaction}, and selects retrieval tools and ranking fusion weights; and the \textbf{Acting Executor} executes the plan through retrieval, candidate merging, and ranking fusion to produce recommendations.
On this basis, we further introduce a \textbf{Dynamic Reflection Tree}.
Combining subsequent user feedback with the execution trace, a Reflection Agent localizes errors to need inference, planning, or previously applied repair strategies, matches them hierarchically against existing generalized error patterns, and generates personalized repair strategies, which are provided to the agents as working memory in subsequent rounds.
As experience accumulates, verified strategies are abstracted into generalized repair strategies, and the tree reorganizes its hierarchy as new error patterns emerge, forming a closed loop of continuous self-evolution over successive recommendation rounds without updating agent parameters~\cite{zhang2026agentic}.

We evaluate EvoRec on local-services advertising data from Kuaishou, a large-scale short-video platform.
EvoRec improves Hit@50 by 15.7\% over the strongest baseline, and reflection improves Hit@50 by 13.1\% for users who received repair strategies.
Controlled ablations further confirm the benefits of adaptive agentic decision-making.
Deployed through nearline generation that keeps LLM inference outside the real-time request path, EvoRec increases advertising revenue by 4.6\% and CTCVR by 3.5\% in a seven-day online A/B test, with a 14.5\% revenue gain among high-intent users.

Our main contributions are summarized as follows:

\begin{itemize}[leftmargin=20pt,topsep=0pt]

    \item \textbf{Unified agentic recommendation.}
    We propose EvoRec, which integrates need inference, recommendation planning, and deterministic tool execution into a unified decision-making process, enabling agents to balance historical interests and inferred needs for each user and to coordinate heterogeneous retrieval tools and ranking models.

    \item \textbf{Dynamic Reflection Tree for self-evolution.}
    We design a Dynamic Reflection Tree that localizes errors to decision stages, verifies personalized repairs by replay, and abstracts verified experience into cross-user strategies while evolving its hierarchy, enabling EvoRec to continuously improve from feedback without updating agent parameters.

    \item \textbf{Industrial deployment and validation.}
    We deploy EvoRec in a production advertising system through nearline generation, keeping LLM inference off the real-time request path. Offline experiments and an online A/B test show that EvoRec outperforms traditional, generative, and LLM-based or agentic recommenders, and ablation studies verify the respective contributions of adaptive decision-making and hierarchical reflection.
\end{itemize}

%
%
%
%
%
%

%

\begin{figure*}[t]
    \centering
    \includegraphics[width=\textwidth]{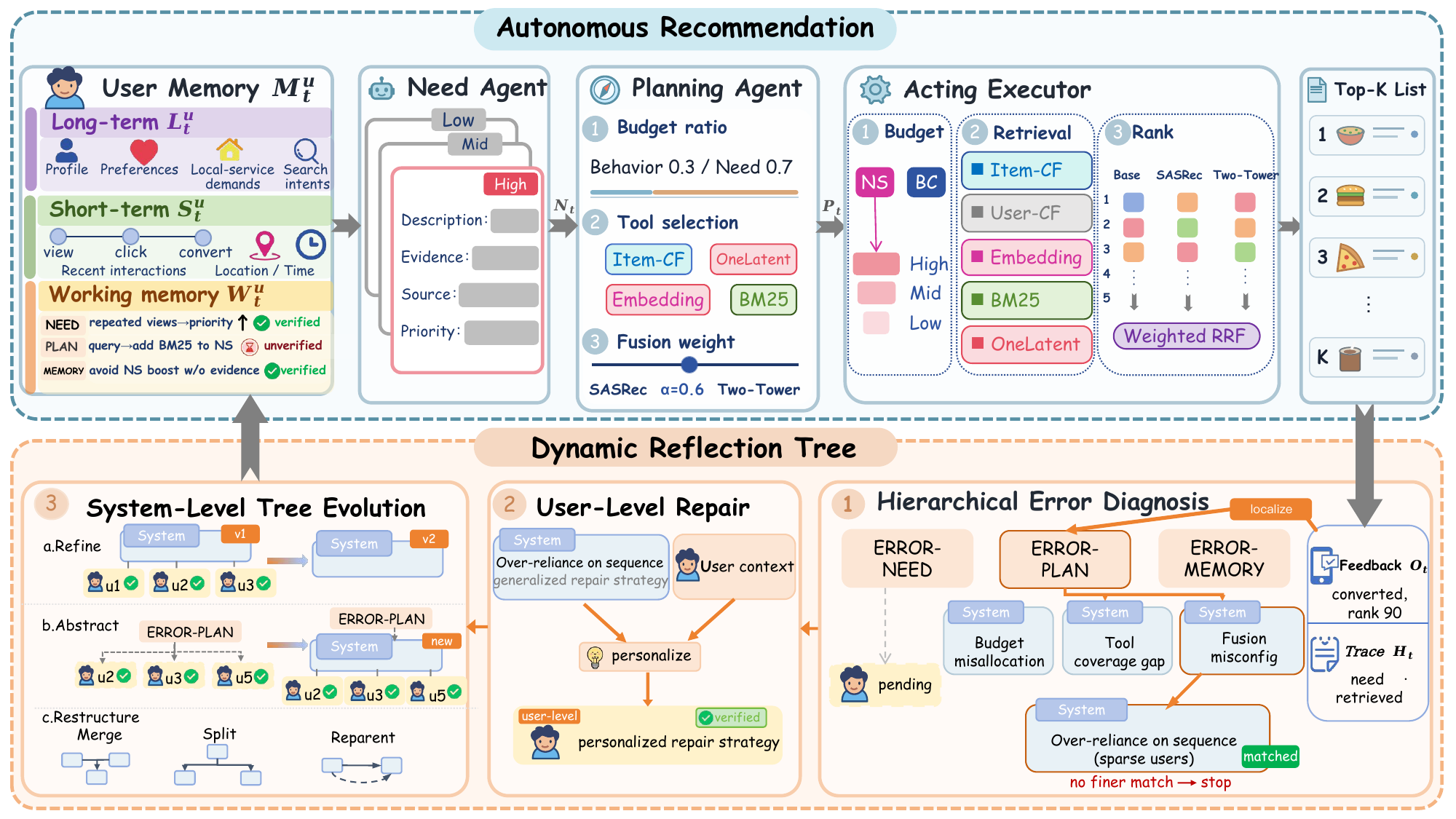}
    \caption{\textbf{Overview of EvoRec.}
    The upper part illustrates autonomous recommendation through need inference, adaptive planning, and deterministic execution.
    The lower part presents the Dynamic Reflection Tree, which performs hierarchical error diagnosis, personalized repair, and system-level evolution.
    Repair strategies are fed back through working memory, enabling continuous self-evolution without parameter updates.}
    \label{fig:framework}
\end{figure*}

\section{Related Work}
\label{sec:related}

\subsection{Large Language Models for Recommendation}
\label{sec:related_llm}

Large language models (LLMs) bring semantic knowledge and richer user modeling to recommender systems. P5~\cite{geng2022recommendation} formulates diverse recommendation tasks as conditional text generation. TALLRec~\cite{bao2023tallrec} and LLaRA~\cite{liao2024llara} adapt pretrained LLMs through recommendation instruction tuning and through the fusion of textual and behavioral representations, respectively. TIGER~\cite{rajput2023recommender} represents items as semantic identifiers for autoregressive generation. OneRec~\cite{deng2025onerec} unifies retrieval and ranking and improves generated recommendations through preference alignment. OpenOneRec~\cite{zhou2025onerec} observes that next-token prediction alone fits the distribution of items exposed by an existing recommender, and therefore uses reinforcement learning to further align recommendations with user preferences. OneRec-Think~\cite{liu2025onerec} incorporates explicit reasoning into recommendation. OneReason~\cite{team2026onereason} combines item semantic alignment, supervised reasoning over user interests, and reinforcement learning to infer latent interests from behavioral evidence. OneLatent~\cite{jiang2026whisperrec} internalizes multi-view reasoning supervision in latent tokens, reducing the cost of explicit chain-of-thought generation while improving recommendation accuracy.
These and related reasoning-enhanced recommenders~\cite{zhang2026interpretable,jiang2026rest} primarily optimize recommendation outputs, with reasoning serving item prediction. Output optimization alone does not specify how user needs guide action selection or how execution feedback informs strategy revision. Agentic recommendation explicitly links analysis, planning, execution, and feedback, extending LLM reasoning from item prediction to planning, executing, and revising recommendation decisions.

\subsection{Agentic Recommendation}
\label{sec:agentic_recommendation}

Recent advances in LLMs have accelerated the development of agentic recommendation, enabling recommender systems to incorporate autonomous reasoning, planning, and tool use. Early frameworks such as InteRecAgent~\cite{huang2025recommender} and MACRec~\cite{wang2024macrec} integrate LLM-based agents with recommendation tools and multi-agent collaboration. Subsequent studies further explore adaptive tool orchestration (ChainRec)~\cite{li2026chainrec}, active information acquisition (RecThinker)~\cite{zhang2026recthinker}, and feedback-driven reasoning optimization (AgenticRec)~\cite{li2026agenticrec}. AgentDR~\cite{yang2026agentdr} combines LLM-based agents with conventional recommendation models, while Cao et al.~\cite{cao2026enhancing} jointly model need prediction and service recommendation. Industrial research has further extended agentic recommendation to large-scale applications. RecGPT~\cite{yi2025recgpt} explores intent-driven recommendation with dynamically evolving user memory; DREAM~\cite{zhang2026dream} introduces an Agentic Meta-Control mechanism based on dual-loop optimization; and AgentX~\cite{lao2026agentx} automates recommendation algorithm development through agent-driven experimentation and iterative refinement. Beyond recommendation, self-reflective agents such as Reflexion~\cite{shinn2023reflexion} and ExpeL~\cite{zhao2024expel} improve task performance by storing verbal reflections or extracted insights from past trials.

Despite these advances, existing agentic recommenders either improve decisions through parameter updates or maintain feedback as global or user-level memory, without localizing errors to specific decision stages or accumulating reusable repair experience across users. General-purpose reflective agents, in turn, operate on tasks with explicit success signals rather than sparse and delayed recommendation feedback. To bridge this gap, we propose EvoRec, a unified agentic recommendation framework that integrates need inference, planning, and tool-based execution. Furthermore, we introduce a Dynamic Reflection Tree that localizes errors to decision stages and organizes repair experience at the user and system levels, transforming recommendation outcomes into actionable experience that guides future recommendation decisions.
%
%
%
%





\section{Methodology}
\label{sec:method}

\subsection{Problem Formulation and Overview}
\label{sec:overview}

Traditional recommendation systems rely on predefined retrieval and ranking pipelines to predict user preferences from historical interactions~\cite{covington2016deep}.
In contrast, EvoRec formulates recommendation as a multi-stage decision-making process that adapts across recommendation rounds.

At round $t$, user $u$'s memory is $M_t^u=(L_t^u, S_t^u, W_t^u)$, comprising the long-term memory $L_t^u$, the short-term memory $S_t^u$, and the working memory $W_t^u$ containing repair strategies derived from reflection.
Given $M_t^u$, the item catalog $\mathcal{I}$, and the recommendation tools $\mathcal{F}$, EvoRec produces a ranked list $Y_t$:
\begin{equation}
\begin{aligned}
    N_t &= f_{\mathrm{need}}\big(M_t^u\big),\\
    P_t &= f_{\mathrm{plan}}\big(M_t^u, N_t\big),\\
    Y_t &= f_{\mathrm{act}}\big(P_t, \mathcal{F}, \mathcal{I}\big),
\end{aligned}
\label{eq:agentic_rec}
\end{equation}
where the Need Agent infers user needs $N_t$, the Planning Agent produces a recommendation plan $P_t$, and the Acting Executor deterministically executes the plan.
Agent parameters remain fixed: while $L_t^u$ and $S_t^u$ are updated with new interactions, reflection-derived experiences influence subsequent decisions through $W_t^u$.

After recommendation, the Reflection Agent uses the decision and execution trace $H_t$ and subsequently observed user feedback $O_t$ to update the Dynamic Reflection Tree:
\begin{equation}
    \mathcal{T}_{t+1} = \mathcal{R}\big(\mathcal{T}_t, H_t, O_t\big).
    \label{eq:reflection}
\end{equation}

The user-specific repair strategies derived from the updated tree form the working memory $W_{t+1}^u$ for the next round.
EvoRec thus comprises \textbf{Autonomous Recommendation} (Section~\ref{sec:agentic_framework}) and \textbf{Reflection-driven Self-Evolution} (Section~\ref{sec:reflection}), as illustrated in Figure~\ref{fig:framework}.

\subsection{Unified Agentic Recommendation Framework}
\label{sec:agentic_framework}

EvoRec organizes autonomous recommendation decision-making into three stages: need inference, planning, and execution, performed by the Need Agent, Planning Agent, and Acting Executor, respectively.

\textbf{Need Agent: User Need Inference.}
The long-term memory $L_t^u$ summarizes the user's profile, preferences, local-service demands, and search intents, while the short-term memory $S_t^u$~\cite{wang2025episodic} records recent ad interactions with their action types, timestamps, and content descriptions, together with the spatiotemporal context of the request.
Combining these memories with the need-related repair strategies in the working memory $W_t^u$ (Section~\ref{sec:reflection}), the Need Agent infers user needs $N_t=\{n_1,\ldots,n_m\}$ following Eq.~(\ref{eq:agentic_rec}).
Each need is represented as a structured record comprising a natural-language description, supporting behavioral evidence, a priority level, and a source type.

\textbf{Planning Agent: Adaptive Recommendation Planning.}
Given the inferred needs $N_t$ and the user memory $M_t^u$, the Planning Agent generates a structured recommendation plan $P_t$ following Eq.~(\ref{eq:agentic_rec}).
To balance historical preferences with inferred user needs, EvoRec adopts two complementary recommendation paths: \textit{Behavior Continuation}, which focuses on historical interests, and \textit{Need Satisfaction}, which targets the inferred user needs.
Guided by plan-related repair strategies from $W_t^u$ (Section~\ref{sec:reflection}), the Planning Agent selects retrieval tools for each path based on the inferred needs and tool capabilities.
It also determines the retrieval budget ratio $w^{\mathrm{NS}}$ assigned to Need Satisfaction, with the remainder assigned to Behavior Continuation, and the fusion weight $\alpha$ of the sequential ranker SASRec relative to the semantic ranker Two-Tower.
These choices are informed by the richness and stability of the user's behavioral history and by the clarity and priority of the inferred needs.
Both ratios are selected from $\{0.1,0.2,\ldots,0.9\}$ to constrain the planning action space; when no need is inferred, $w^{\mathrm{NS}}$ is set to $0$ and the entire budget is assigned to Behavior Continuation.

\textbf{Acting Executor: Deterministic Plan Execution.}
The Acting Executor carries out $P_t$ through a unified callable interface to retrieval tools and rankers.
The retrieval tools cover complementary signals: Item-CF~\cite{sarwar2001item} and User-CF~\cite{ferrari2019we} exploit item associations and similar-user preferences, embedding-based retrieval and BM25~\cite{robertson2009probabilistic} match items with user interests and inferred needs at the semantic and lexical levels, and OneLatent~\cite{jiang2026whisperrec} leverages latent reasoning representations for recommendation.
Additional retrieval methods~\cite{zhang2025video,zhang2025enhancing} could be integrated as tools through adapters that map their outputs to ranked item candidates.
The executor assigns budgets $w^{\mathrm{NS}}B$ and $(1-w^{\mathrm{NS}})B$ to Need Satisfaction and Behavior Continuation, where $B$ is the total retrieval budget, and further allocates $w^{\mathrm{NS}}B\,\rho(n_k)/\sum_{l}\rho(n_l)$ to each inferred need $n_k$, where $\rho(\cdot)$ maps the high, medium, and low priority levels to fixed weights.
It then invokes the selected tools accordingly; when the tools return fewer candidates than their budgets, a predefined backfill procedure fills the remaining budget.
The retrieved lists are merged by snake-order interleaving, which preserves the internal ranking of each tool, and deduplicated to form the candidate set $\mathcal{C}_t$ and its Base ranking $Y_t^{\mathrm{base}}$.
The rankers SASRec~\cite{kang2018self} and Two-Tower~\cite{yi2019sampling} then rank $\mathcal{C}_t$, and the three rankings are combined by weighted reciprocal rank fusion (RRF)~\cite{cormack2009reciprocal}:
\begin{equation}
    s_t(i)=
    \sum_{m\in\{\mathrm{base},\mathrm{seq},\mathrm{tower}\}}
    \frac{w_{t,m}}{c+r_{t,m}(i)},
    \qquad i\in\mathcal{C}_t,
    \label{eq:acting_rrf}
\end{equation}
where $\mathrm{base}$, $\mathrm{seq}$, and $\mathrm{tower}$ denote the Base, SASRec, and Two-Tower rankings, respectively; $r_{t,m}(i)$ is the rank of item $i$ in ranking $m$, and $c$ is a rank-smoothing constant.
The Base ranking receives a fixed weight, while the plan splits a unit weight between the two learned rankers, with $w_{t,\mathrm{seq}}=\alpha$ and $w_{t,\mathrm{tower}}=1-\alpha$.
Sorting candidates in descending order of $s_t(i)$ yields the final list $Y_t$.
Throughout execution, the executor records the inferred needs, plan, tool calls, budgets, candidate provenance, and final rankings, forming the trace $H_t$ for reflection.

\subsection{Dynamic Reflection Tree for Self-Evolution}
\label{sec:reflection}

Each recommendation round yields an outcome, yet final feedback alone neither reveals where a decision went wrong nor makes past experience reusable. We therefore introduce a Dynamic Reflection Tree that organizes reflective experience at the system and user levels. Formally, the tree is denoted as $\mathcal{T}_t=(\mathcal{V}^{\mathrm{sys}}_t\cup\mathcal{V}^{\mathrm{usr}}_t,\mathcal{E}_t)$. Each system-level node $v=(d_v,r_v)$ consists of an error description $d_v$ and a generalized repair strategy $r_v$; parent--child relations among system-level nodes form an error hierarchy whose top level comprises \texttt{ERROR-NEED} for need-inference errors, \texttt{ERROR-PLAN} for planning errors, and \texttt{ERROR-MEMORY} for errors induced by repair strategies in the working memory. Each user-level node $e=(u_e,d_e,r_e,\sigma_e)$ consists of the user $u_e$, a description of the diagnostic evidence $d_e$, a personalized repair strategy $r_e$, and a verification status $\sigma_e\in\{\textit{unverified},\textit{verified},\textit{retracted}\}$, and is linked to its matched system-level nodes. The Reflection Agent diagnoses errors and generates personalized repair strategies, and, as user-level experiences accumulate, continuously refines generalized repair strategies and evolves the tree structure.

\textbf{Hierarchical Error Diagnosis.}
Given the trace $H_t$ and subsequent feedback $O_t$, the Reflection Agent extracts structured diagnostic signals for each feedback-associated item, including evidence for the feedback-indicated need in the long- and short-term memory, whether and with what priority the Need Agent inferred this need, retrieval hits of the item across paths and tools, its rank in $Y_t$, and the working-memory nodes provided to each agent. Based on these signals, the Reflection Agent localizes potential errors to the three top-level nodes of the tree. If the long- or short-term memory contains evidence for the feedback-indicated need but the Need Agent fails to infer it or assigns an inappropriate priority, the error is localized to \texttt{ERROR-NEED}; if such evidence is absent, need inference is not held responsible. If the inferred needs are adequate but the item is not retrieved or falls outside the top-$K$ positions of $Y_t$, and the Reflection Agent attributes this outcome to the budget ratio, tool selection, or fusion weight based on the per-path and per-tool retrieval results in the trace, the error is localized to \texttt{ERROR-PLAN}; such attributions are subsequently checked by replay verification. If re-executing the decision without the working-memory nodes provided to the corresponding agent eliminates a need or plan error, indicating that previously generated repair strategies misled the decision, the error is localized to \texttt{ERROR-MEMORY} instead of \texttt{ERROR-NEED} or \texttt{ERROR-PLAN}. Multiple errors with independent supporting evidence may be identified, while cases satisfying none of these conditions are not attributed.

The Reflection Agent then invokes the Tree Tool to perform hierarchical matching over system-level nodes. The Tree Tool first returns the immediate children of each identified top-level node in a batch, and the Reflection Agent jointly evaluates these candidates against the diagnostic evidence to determine whether they match the current error. For each selected node, the Tree Tool returns its immediate children, allowing the Reflection Agent to determine whether a more specific match exists. This process iterates along the selected branches until no more specific node matches, yielding a set of matched system-level nodes; an error with no matching child under its top-level node is treated as unmatched.

\textbf{User-Level Repair and Adaptation.}
For matched errors, the Reflection Agent combines the generalized repair strategies of the matched system-level nodes with the current user context to generate a personalized repair strategy, and records the diagnostic evidence and the strategy as a user-level error node linked to these system-level nodes. For unmatched errors, it generates a personalized repair strategy directly from the diagnostic evidence and user context, and links the resulting user-level node to the corresponding top-level node as a pending node awaiting abstraction. Repair strategies specify the condition, direction, and approximate magnitude of adjustments to need inference and planning rather than which items to recommend, preventing repairs from degenerating into memorization of specific feedback items and allowing them to generalize to subsequent recommendations.

To verify a personalized repair strategy, the Reflection Agent injects it into the original recommendation context, re-executes the pipeline, and checks whether the diagnosed error is mitigated with respect to the observed feedback~\cite{zhu2025llm}. The node is marked as \textit{verified} if the error is mitigated and \textit{unverified} otherwise. Node status is continuously updated with subsequent feedback: a strategy implicated in \texttt{ERROR-MEMORY} is marked as \textit{retracted}, while an unverified node is promoted to \textit{verified} once its strategy mitigates a recurrence of the diagnosed error in a later round.

Personalized repair strategies are applied to subsequent recommendations through working memory rather than by modifying the execution pipeline~\cite{shinn2023reflexion,ouyang2026reasoningbank}. The working memory of user $u$ consists of the user's non-retracted user-level nodes, $W_t^u=\{e\in\mathcal{V}^{\mathrm{usr}}_t\mid u_e=u,\ \sigma_e\neq\textit{retracted}\}$. In subsequent recommendations, the repair strategies in $W_t^u$ are inserted into the prompts of the Need Agent or the Planning Agent according to their top-level category: strategies under \texttt{ERROR-NEED} direct the Need Agent to recover missed needs and revise need priorities; strategies under \texttt{ERROR-PLAN} direct the Planning Agent to adjust $w^{\mathrm{NS}}$, tool selection, and $\alpha$, with the adjusted ratios still selected from their grids; and strategies under \texttt{ERROR-MEMORY} are provided to the agent whose earlier strategy was retracted, cautioning it against similar adjustments. Consequently, the self-evolution of EvoRec takes place entirely at inference time, without updating the parameters of any agent.

\textbf{System-Level Tree Evolution.}
As verified user-level nodes accumulate, the Reflection Agent consolidates them into system-level knowledge. For nodes linked to existing system-level nodes, it extracts patterns shared across users to refine the corresponding generalized repair strategies. For pending nodes, once a sufficient number of verified ones has accumulated under the same top-level node, the Reflection Agent analyzes them collectively and abstracts them into new system-level nodes with generalized repair strategies under that top-level node. As error patterns expand, the Reflection Agent further reorganizes the hierarchy: \textit{Merge} consolidates redundant nodes describing the same error mechanism, \textit{Split} separates a node covering distinct mechanisms into more specific children, and \textit{Reparent} corrects inappropriate parent--child relations. Each structural operation is checked for error-mechanism consistency and hierarchical validity before being applied.

\begin{figure}[t]
  \centering  
\includegraphics[width=\linewidth,height=0.8\textheight,keepaspectratio]{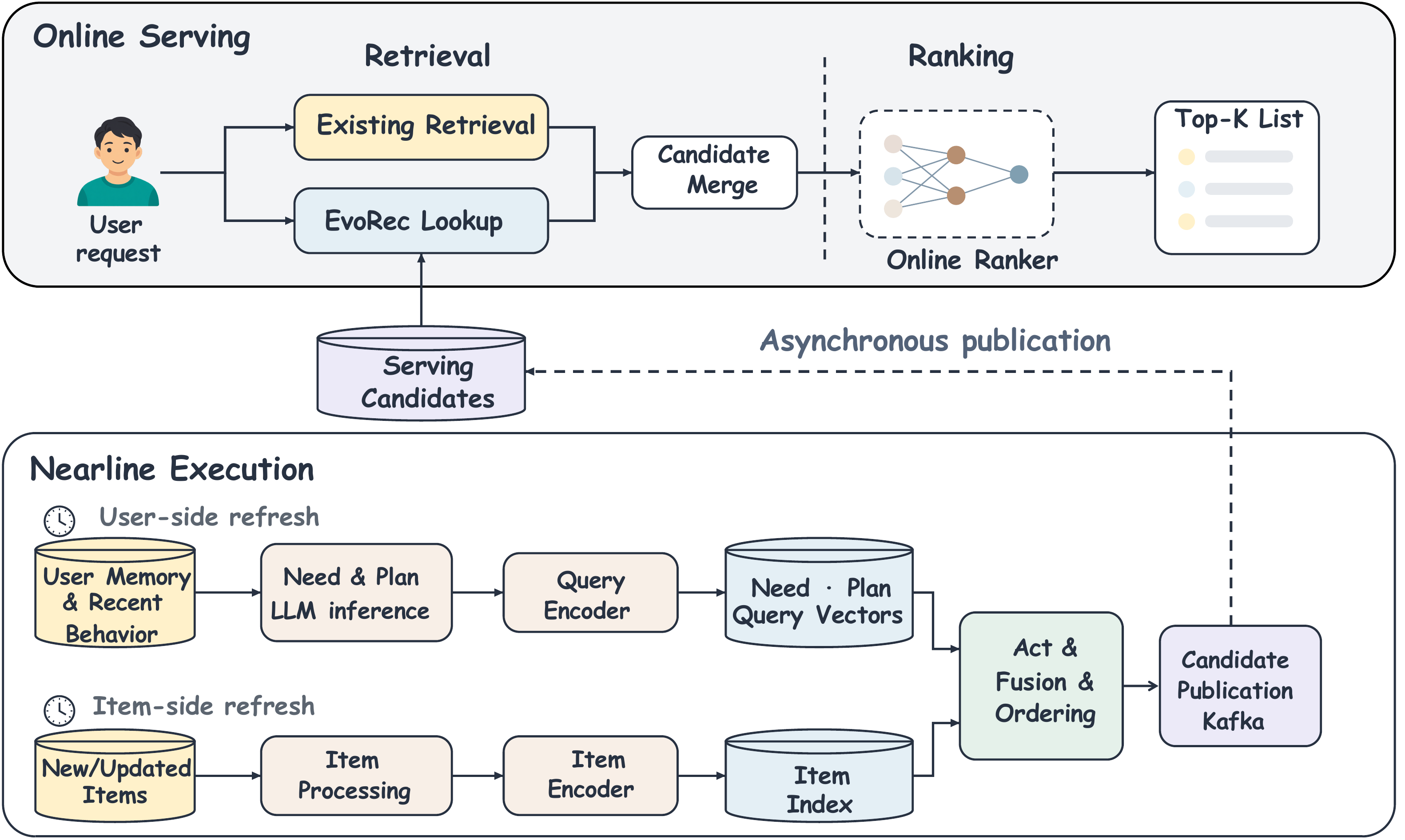}
  \caption{The online deployment architecture of EvoRec.}
  \label{fig:deployment}
\end{figure}

\begin{table*}[t]
    \centering
    \caption{Performance comparison of recommendation methods.
    Hit@$K$ is reported as a percentage.}
    \label{tab:main_results}

    \begingroup
    \normalsize
    \setlength{\tabcolsep}{3pt}
    \renewcommand{\arraystretch}{1.15}
    \newcommand{\tabNA}{\textemdash}

    \begin{tabular*}{\textwidth}{
        @{\extracolsep{\fill}}l*{11}{c}@{}
    }
        \toprule

        \multirow{2}{*}{\textbf{Method}}
        & \multicolumn{6}{c}{\textbf{Hit@$K$} $\uparrow$}
        & \multicolumn{3}{c}{\textbf{NDCG@$K$} $\uparrow$}
        & \multicolumn{2}{c}{\textbf{ILD@$K$} $\uparrow$} \\

        \cmidrule(lr){2-7}
        \cmidrule(lr){8-10}
        \cmidrule(lr){11-12}

        & \textbf{@1}
        & \textbf{@5}
        & \textbf{@10}
        & \textbf{@20}
        & \textbf{@50}
        & \textbf{@100}
        & \textbf{@5}
        & \textbf{@10}
        & \textbf{@20}
        & \textbf{@50}
        & \textbf{@100} \\

        \midrule

        SASRec
        & 1.35 & 4.76 & 7.13 & 10.77 & 16.68 & 22.06
        & 0.0299 & 0.0373 & 0.0462
        & 0.5781 & 0.5899 \\

        UltraGCN
        & 0.91 & 3.19 & 4.78 & 8.04 & 13.32 & 17.67
        & 0.0201 & 0.0250 & 0.0330
        & 0.6369 & 0.6441 \\

        TIGER
        & 1.32 & 4.73 & 7.09 & 10.86 & 16.97 & 22.85
        & 0.0293 & 0.0371 & 0.0463
        & 0.4231 & 0.4322 \\

        OneRec
        & 1.42 & 5.01 & 7.51 & 11.61 & 18.26 & 24.82
        & 0.0315 & 0.0393 & 0.0493
        & 0.5181 & 0.5499 \\

        LLMRank
        & 0.92 & 3.86 & 6.41 & 10.12 & 17.34 & 25.72
        & 0.0241 & 0.0329 & 0.0428
        & 0.5376 & 0.5412 \\
        
        MACRec
        & 1.01 & 4.12 & 6.79 & 10.86 & 18.21 & 25.72
        & 0.0262 & 0.0351 & 0.0461
        & 0.5358 & 0.5412 \\
        
        ToolRec
        & 1.73 & 6.21 & 9.38 & 13.84 & 20.47 & 26.61
        & 0.0398 & 0.0497 & 0.0602
        & 0.5287 & 0.5396 \\
        
        InteRecAgent
        & 1.87 & 6.53 & 9.76 & 14.29 & 21.16 & 27.42
        & 0.0415 & 0.0521 & 0.0631
        & 0.5302 & 0.5411 \\

        \midrule

        EvoRec w/o Reflection
        & 2.33 & 7.52 & 11.37 & 16.23 & 23.90 & 30.38
        & 0.0479 & 0.0600 & 0.0720
        & 0.5339 & 0.5438 \\

        \rowcolor{black!6}
        \textbf{EvoRec}
        & \textbf{2.37} & \textbf{7.66} & \textbf{11.61}
        & \textbf{16.62} & \textbf{24.48} & \textbf{31.06}
        & \textbf{0.0488} & \textbf{0.0612} & \textbf{0.0736}
        & 0.5364 & 0.5464 \\

        \bottomrule
    \end{tabular*}
    \endgroup
\end{table*}

\subsection{Industrial Deployment}
\label{sec:industrial_deployment}

To integrate EvoRec into the existing recommendation pipeline, we adopt nearline generation and online consumption, keeping LLM inference outside the real-time request path. As shown in Figure~\ref{fig:deployment}, the nearline user pipeline performs need inference and resource planning through Need and Plan, respectively, using user memory and recent behavior. Retrieval queries are encoded in advance, and the resulting needs, plans, and query vectors are stored for reuse by Act. Because user needs and item supply evolve at different rates, user computation and item updates run independently, as depicted by the two branches in the lower part of the figure. The item branch preprocesses and encodes new or updated items to maintain the retrieval index. Act allocates retrieval resources according to the stored plan and uses the saved needs and query vectors to retrieve candidates from the current index, followed by deduplication, fusion, and ordering. Provided that the user-side results remain applicable and the query and item encodings remain compatible, an index update can be followed by another Act execution to incorporate newly available items without repeating need inference, planning, or query encoding. This separation amortizes the cost of user-side reasoning across multiple candidate updates while allowing retrieval to reflect changes in item supply.

The resulting candidate lists are published asynchronously through Kafka for each user, connecting nearline execution to online serving in the upper part of Figure~\ref{fig:deployment}. EvoRec operates in parallel with the existing retrieval routes as an additional source of candidates. Upon request, the serving system retrieves the user's precomputed EvoRec candidates and combines them with candidates from the existing routes. Candidates from all routes then compete jointly in the same fine-ranking stage, where the online ranker uses the current request context to produce the final recommendation list. While Act orders candidates for nearline publication, their final selection and ordering are determined by the shared online ranker. LLM decisions therefore influence online recommendations through precomputed candidates, without requiring synchronous need inference, resource planning, or query encoding during request processing.

Reflection is restricted to high-intent users with advertising inquiry or search intent, accounting for 11.3\% of active users. Their richer feedback and clearer needs enable more reliable and cost-effective reflection. Reflection runs offline and updates user-level experience upon new clicks or conversions. The resulting working memory is stored and made available to nearline Need and Plan upon their next refresh. This feedback loop allows observed user responses to inform subsequent need inference and resource planning without adding computation to online requests.


\section{Experiments}
\label{sec:experiments}

\subsection{Experimental Setup}
\label{sec:setup}

\textbf{Dataset.}
We construct a proprietary dataset from logs of local-services recommendation ads on Kuaishou, a large-scale short-video platform, comprising user and item profiles and four interaction types: impressions, 3-second views, clicks, and conversions.
Given all information available by the end of day $D$, the task is to rank candidate items for each user, with items converted by the user on day $D+1$ serving as ground truth.
The evaluation set consists of 28,227 users, all sharing a candidate pool of 355,163 items from the snapshot on day $D$.
The dataset is anonymized, with de-identified user IDs and no personally identifiable information retained.

\textbf{Evaluation Protocol.}
EvoRec is evaluated after seven rounds of reflection over the preceding week, starting from a tree with only its three top-level nodes and empty working memory.
On each day $d\in\{D-6,\ldots,D\}$, EvoRec generates recommendations using user memory available by the end of day $d-1$, and after the day ends, clicks and conversions from evaluation users drive reflection.
On day $D+1$, EvoRec uses the working memory accumulated through day $D$, and all methods use only information available by the end of day $D$ and are evaluated against conversions observed on day $D+1$.
Section~\ref{sec:analysis} extends this protocol to subsequent days to examine how reflection evolves.

\textbf{Baselines.}
We compare EvoRec with three groups of representative baselines:
(1) \textit{traditional recommenders}, including SASRec~\cite{kang2018self} and UltraGCN~\cite{mao2021ultragcn};
(2) \textit{generative recommenders}, including TIGER~\cite{rajput2023recommender} and OneRec~\cite{deng2025onerec};
and (3) \textit{LLM-based and agentic recommenders}, including LLMRank~\cite{hou2024large}, MACRec~\cite{wang2024macrec}, ToolRec~\cite{zhao2024let}, and InteRecAgent~\cite{huang2025recommender}, all using Qwen3-32B~\cite{yang2025qwen3}.
LLMRank and MACRec, which are designed to rank given candidates, rerank the top-100 candidates of the Base ranking $Y_t^{\mathrm{base}}$ produced by EvoRec's retrieval.
ToolRec and InteRecAgent, which natively invoke retrieval tools, are given the same user memory, retrieval tools, rankers, and retrieval budget as EvoRec, so that they differ from EvoRec only in how recommendation decisions are made.
Detailed baseline descriptions are provided in Appendix~\ref{app:baselines}.

We additionally include \textit{Base}, our agentic recommendation configuration without the proposed self-evolution mechanism, to assess the contribution of reflection-driven experience reuse. EvoRec (Full System) further incorporates the Dynamic Reflection Tree for hierarchical error matching, personalized repair, and experience evolution.

\textbf{Evaluation Metrics.}
We evaluate recommendation accuracy using Hit@$K$ and NDCG@$K$, and semantic diversity using ILD@$K$.
All metrics are computed over the top-$K$ recommendations,
with cutoffs specified in the tables.
Accuracy metrics are averaged across evaluation users.
ILD@$K$ measures the mean pairwise cosine distance between
item embeddings and is averaged over users with at least
two valid embeddings.
Hit is reported as a percentage.

\textbf{Implementation Details.}
The Need Agent and Planning Agent use Qwen3-4B~\cite{yang2025qwen3}, while the Reflection Agent uses Qwen3-32B~\cite{yang2025qwen3}. The total retrieval budget is set to 3,000 candidates per user. Further implementation details are provided in Appendix~\ref{app:implementation}.


\subsection{Main Results}
\label{sec:main-results}

As shown in Table~\ref{tab:main_results}, EvoRec achieves the best results across all accuracy metrics, improving Hit@10 by 19.0\%, Hit@50 by 15.7\%, and NDCG@10 by 17.5\% over InteRecAgent, the strongest baseline.
Even without reflection, EvoRec surpasses InteRecAgent by 12.9\% in Hit@50 despite identical tools, rankers, and retrieval budget and a smaller decision model (Qwen3-4B vs.\ Qwen3-32B), indicating that its gains stem from the decision mechanism rather than tool aggregation; in contrast, LLMRank and MACRec, which only rerank the top-100 Base candidates, fall below the Base ranking itself (Hit@10 of 8.97\%, Table~\ref{tab:ranking_fusion_1}).
Reflection further improves Hit@50 by 2.4\%, with gains concentrated on the 18.5\% of users who received working memory, on whom Hit@50 increases by 13.1\% (Table~\ref{tab:reflection_ablation}).
EvoRec also maintains diversity comparable to OneRec and the LLM-based baselines, while SASRec and UltraGCN attain higher ILD at the cost of substantially lower accuracy.

\subsection{Ablation Study}
\label{sec:ablation}
\textbf{Agentic vs.\ Static Decisions.}
Table~\ref{tab:agentic_ablation} tests whether EvoRec's gains stem from adaptive decisions rather than tool aggregation. All variants have access to the same tools, total retrieval budget, and ranking models, with reflection disabled throughout.
\textit{w/o Need Agent} removes need inference and the Need Satisfaction path, while the Planning Agent still adapts tool budgets and fusion weights within Behavior Continuation.
\textit{w/o Planning Agent} retains the inferred needs and both paths but fixes path weights, tool budgets, and fusion weights.
\textit{Static Ensemble} removes both need inference and adaptive planning, using only Behavior Continuation with fixed budgets and fusion weights.
Fixed configurations follow the average allocations under EvoRec.
Static Ensemble reduces Hit@50 from 23.90\% to 20.69\% ($-$13.4\%), showing that EvoRec's gains cannot be explained by tool aggregation alone.
Removing the Need Agent reduces Hit@50 by 7.2\%, more than at Hit@10 (5.8\%), suggesting that inferred needs help retrieve items less accessible from behavioral history, while removing the Planning Agent yields a smaller decline (3.1\%), indicating that adaptive allocation remains beneficial even with inferred needs.

\begin{table}[t]
    \centering
    \caption{Agentic vs.\ static decisions with the same available tools, retrieval budget, and ranking models. Hit@$K$ in \%. $\Delta$: relative change in Hit@50 w.r.t.\ EvoRec w/o Reflection.}
    \label{tab:agentic_ablation}

    \begingroup
    \small
    \setlength{\tabcolsep}{2pt}
    \renewcommand{\arraystretch}{1.1}

    \begin{tabular*}{\columnwidth}{
        @{\extracolsep{\fill}}l*{6}{c}@{}
    }
        \toprule

        \multirow{2}{*}{\textbf{Variant}}
        & \multicolumn{3}{c}{\textbf{Hit@$K$} $\uparrow$}
        & \multicolumn{2}{c}{\textbf{NDCG@$K$} $\uparrow$}
        & \multirow{2}{*}{$\boldsymbol{\Delta}$} \\

        \cmidrule(lr){2-4}
        \cmidrule(lr){5-6}

        & \textbf{@5}
        & \textbf{@10}
        & \textbf{@50}
        & \textbf{@5}
        & \textbf{@10}
        & \\

        \midrule

        Static Ensemble
        & 6.85 & 9.42 & 20.69
        & 0.0428 & 0.0516
        & $-$13.4\% \\

        w/o Need Agent
        & 7.08 & 10.71 & 22.18
        & 0.0452 & 0.0566
        & $-$7.2\% \\

        w/o Planning Agent
        & \underline{7.27} & \underline{11.03} & \underline{23.16}
        & \underline{0.0465} & \underline{0.0584}
        & $-$3.1\% \\

        \midrule

        \rowcolor{black!6}
        \textbf{EvoRec w/o Reflection}
        & \textbf{7.52} & \textbf{11.37} & \textbf{23.90}
        & \textbf{0.0479} & \textbf{0.0600}
        & \textemdash \\

        \bottomrule
    \end{tabular*}
    \endgroup
\end{table}

\begin{table}[t]
    \centering
    \caption{Ablation of retrieval tools on retrieval-only results. Hit@$K$ in \%; Hit@3000 equals the coverage of the candidate set. $\Delta$: relative change in Hit@50 w.r.t.\ all tools.}
    \label{tab:recall_ablation_1}

    \begingroup
    \small
    \setlength{\tabcolsep}{2pt}
    \renewcommand{\arraystretch}{1.1}

    \begin{tabular*}{\columnwidth}{
        @{\extracolsep{\fill}}l*{5}{c}@{}
    }
        \toprule

        \multirow{2}{*}{\textbf{Configuration}}
        & \multicolumn{4}{c}{\textbf{Hit@$K$} $\uparrow$}
        & \multirow{2}{*}{$\boldsymbol{\Delta}$} \\

        \cmidrule(lr){2-5}

        & \textbf{@50}
        & \textbf{@100}
        & \textbf{@500}
        & \textbf{@3000}
        & \\

        \midrule

        \rowcolor{black!6}
        \textbf{All tools}
        & \textbf{20.02}
        & \textbf{25.72}
        & \textbf{37.68}
        & \textbf{52.55}
        & \textemdash \\

        \midrule

        w/o Item-CF
        & 17.59
        & 22.05
        & 32.15
        & 46.94
        & $-$12.1\% \\

        w/o User-CF
        & 18.01
        & 23.85
        & 35.07
        & \underline{51.85}
        & $-$10.0\% \\

        \midrule

        w/o Embedding
        & 19.18
        & 24.47
        & 35.49
        & 48.26
        & $-$4.2\% \\

        w/o BM25
        & \underline{19.63}
        & \underline{25.18}
        & \underline{36.77}
        & 51.09
        & $-$1.9\% \\

        \midrule

        w/o OneLatent
        & 19.41
        & 24.79
        & 36.58
        & 51.31
        & $-$3.0\% \\

        \bottomrule
    \end{tabular*}
    \endgroup
\end{table}

\textbf{Retrieval Tool Collaboration.}
Table~\ref{tab:recall_ablation_1} evaluates each retrieval tool without downstream ranking.
Removing any tool degrades retrieval at all cutoffs, most notably Item-CF and User-CF ($-$12.1\% and $-$10.0\% in Hit@50), while OneLatent and BM25 contribute smaller but non-redundant gains ($-$3.0\% and $-$1.9\%).
Removing embedding-based retrieval reduces Hit@50 by 4.2\% but Hit@3000 by 8.2\%, indicating that it mainly extends coverage at larger depths.
These results show that the tools provide complementary signals, supporting the Planning Agent's coordination across them.

\begin{table}[t]
    \centering
    \caption{Comparison of ranking strategies. Base: snake-order merge of retrieved candidates; +: RRF fusion with Base; TT: Two-Tower; fixed/agentic: balanced preset for all users vs.\ per-user presets. Hit@$K$ in \%. $\Delta$: relative change in Hit@50 w.r.t.\ the agentic fusion.}
    \label{tab:ranking_fusion_1}

    \begingroup
    \small
    \setlength{\tabcolsep}{2pt}
    \renewcommand{\arraystretch}{1.1}

    \begin{tabular*}{\columnwidth}{
        @{\extracolsep{\fill}}l*{6}{c}@{}
    }
        \toprule

        \multirow{2}{*}{\textbf{Strategy}}
        & \multicolumn{3}{c}{\textbf{Hit@$K$} $\uparrow$}
        & \multicolumn{2}{c}{\textbf{NDCG@$K$} $\uparrow$}
        & \multirow{2}{*}{$\boldsymbol{\Delta}$} \\

        \cmidrule(lr){2-4}
        \cmidrule(lr){5-6}

        & \textbf{@5}
        & \textbf{@10}
        & \textbf{@50}
        & \textbf{@5}
        & \textbf{@10}
        & \\

        \midrule

        Base
        & 5.77 & 8.97 & 20.02
        & 0.0365 & 0.0465
        & $-$16.2\% \\

        SASRec
        & 4.79 & 7.38 & 17.13
        & 0.0302 & 0.0383
        & $-$28.3\% \\

        Two-Tower
        & 4.16 & 6.18 & 13.82
        & 0.0262 & 0.0325
        & $-$42.2\% \\

        \midrule

        + SASRec
        & 6.22 & 9.85 & 21.51
        & 0.0389 & 0.0503
        & $-$10.0\% \\

        + SASRec + TT (fixed)
        & \underline{7.26}
        & \underline{11.10}
        & \underline{23.28}
        & \underline{0.0461}
        & \underline{0.0582}
        & $-2.6\%$ \\

        \rowcolor{black!6}
        \textbf{+ SASRec + TT (agentic)}
        & \textbf{7.52} & \textbf{11.37} & \textbf{23.90}
        & \textbf{0.0479} & \textbf{0.0600}
        & \textemdash \\

        \bottomrule
    \end{tabular*}
    \endgroup
\end{table}

\textbf{Ranking Signal Fusion.}
Table~\ref{tab:ranking_fusion_1} compares ranking strategies over the same retrieved candidates.
Although SASRec and Two-Tower individually underperform Base, fusing them with Base raises Hit@50 from 20.02\% to 23.28\% under fixed weights, showing that even weaker standalone rankers provide complementary evidence.
Letting the Planning Agent set fusion weights per user further increases Hit@50 to 23.90\%, indicating that adapting the relative contributions of rankers to individual users improves ranking.

\begin{figure}[!t]
  \centering  
\includegraphics[width=\linewidth,height=0.6\textheight,keepaspectratio]{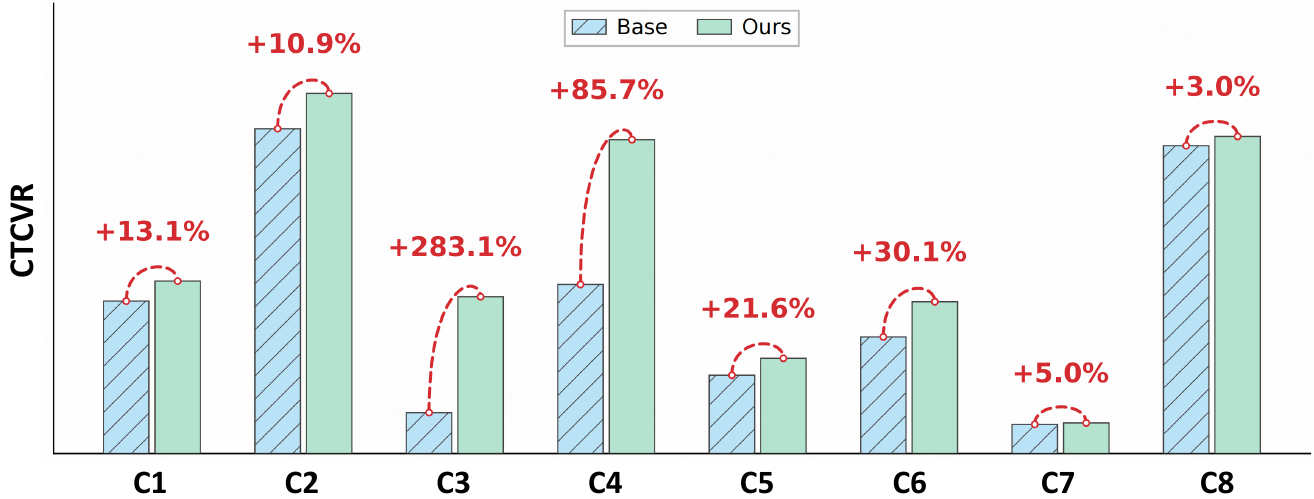}
  \caption{Online CTCVR across eight industry categories, anonymized as C1--C8 and ordered by descending revenue.}
  \label{online_by_industry}
\end{figure}

\begin{table}[t]
    \centering
    \caption{Ablation of the Dynamic Reflection Tree on 5,216 users (18.5\% of evaluation users) who received working memory during seven rounds of reflection, evaluated on day $D+1$; for the remaining users, all variants yield identical recommendations. Hit@$K$ in \%. $\Delta$: relative change in Hit@50 w.r.t.\ EvoRec.}
    \label{tab:reflection_ablation}

    \begingroup
    \small
    \setlength{\tabcolsep}{2pt}
    \renewcommand{\arraystretch}{1.1}

    \begin{tabular*}{\columnwidth}{
        @{\extracolsep{\fill}}l*{6}{c}@{}
    }
        \toprule

        \multirow{2}{*}{\textbf{Variant}}
        & \multicolumn{3}{c}{\textbf{Hit@$K$} $\uparrow$}
        & \multicolumn{2}{c}{\textbf{NDCG@$K$} $\uparrow$}
        & \multirow{2}{*}{$\boldsymbol{\Delta}$} \\

        \cmidrule(lr){2-4}
        \cmidrule(lr){5-6}

        & \textbf{@5}
        & \textbf{@10}
        & \textbf{@50}
        & \textbf{@5}
        & \textbf{@10}
        & \\

        \midrule

        \rowcolor{black!6}
        \textbf{EvoRec}
        & \textbf{8.38}
        & \textbf{13.25}
        & \textbf{27.42}
        & \textbf{0.0529}
        & \textbf{0.0682}
        & \textemdash \\

        w/o Reflection
        & 7.61
        & 11.94
        & 24.25
        & 0.0481
        & 0.0617
        & $-$11.6\% \\

        \midrule

        Flat Reflection
        & 7.96
        & 12.48
        & 25.61
        & 0.0502
        & 0.0645
        & $-$6.6\% \\

        w/o Personalization
        & 7.86
        & 12.35
        & 25.25
        & 0.0496
        & 0.0636
        & $-$7.9\% \\

        w/o Tree Evolution
        & 8.15
        & 12.90
        & 26.57
        & 0.0516
        & 0.0664
        & $-$3.1\% \\

        \midrule

        w/o \texttt{ERROR-MEMORY}
        & \underline{8.26}
        & \underline{13.04}
        & \underline{26.96}
        & \underline{0.0522}
        & \underline{0.0672}
        & $-$1.7\% \\

        w/o Verification
        & 8.03
        & 12.65
        & 25.96
        & 0.0507
        & 0.0652
        & $-$5.3\% \\
        \bottomrule
    \end{tabular*}
    \endgroup
\end{table}

\textbf{Reflection Mechanism.}
Table~\ref{tab:reflection_ablation} evaluates the Dynamic Reflection Tree on the 5,216 users (18.5\% of evaluation users) who received working memory during seven rounds of reflection, with each variant modifying a single component.
On these users, EvoRec improves Hit@50 from 24.25\% to 27.42\% and NDCG@10 from 0.0617 to 0.0682 over w/o Reflection, corresponding to relative gains of 13.1\% and 10.5\%, which translate into the 2.4\% gain in Hit@50 on all users in Table~\ref{tab:main_results}.
Regarding the knowledge structure, w/o Personalization applies generalized strategies without user-specific instantiation, reducing Hit@50 by 7.9\%, while Flat Reflection generates repair strategies without system-level nodes, causing a 6.6\% decline, highlighting the contributions of both shared error knowledge and personalization.
Regarding reliability, w/o Verification admits unverified strategies into system-level abstraction, reducing Hit@50 by 5.3\%, indicating that verification is necessary for reliable generalization.
Disabling tree evolution (while retaining node creation) and retraction via \texttt{ERROR-MEMORY} yield smaller declines of 3.1\% and 1.7\%, respectively; their cumulative effects are further examined across days in Section~\ref{sec:analysis}.

\subsection{Online Evaluation}
\label{sec:online}






We conducted a seven-day online A/B test from August 28 to September 3, 2026, for lead-generation ads in Kuaishou's
local-services recommendation system.
Users were randomly assigned via user-ID hashing, with the control and treatment groups each accounting for approximately
5\% of eligible traffic. Here, high-intent users are defined as those with a non-empty advertising inquiry intent field or search intent field. These fields are derived from users' inquiry messages in
advertising contexts and their search queries, respectively.
We report only relative changes from the control group
to protect commercially sensitive information.
Revenue denotes advertising revenue billed by the platform
during the experiment.

As shown in Table~\ref{tab:online_ab}, EvoRec increases impressions by 0.8\%, advertising revenue by 4.6\%,
and CTCVR by 3.5\% within its deployment scope.
Among high-intent users, advertising revenue increases
by 14.5\% and CTCVR improves by 5.6\%, compared with
a 5.7\% increase in cost.
These results show improvements in both revenue and CTCVR,
with larger relative gains among high-intent users.
For high-intent users, inquiry messages or search queries
provide explicit evidence of user needs.
The larger gains in this cohort suggest that EvoRec's
autonomous recommendation process can translate such
evidence into effective recommendation actions through
need inference, planning, and tool execution.

\begin{figure}[!t]
\centering
\begin{tikzpicture}
\begin{axis}[
    scale only axis,
    width=0.85\linewidth, height=2.8cm,
    ymin=-0.4, ymax=3.3, xmin=-0.4, xmax=9.4,
    xtick={0,...,9},
    xticklabels={$D{-}5$,$D{-}4$,$D{-}3$,$D{-}2$,$D{-}1$,$D$,$D{+}1$,$D{+}2$,$D{+}3$,$D{+}4$},
    tick label style={font=\scriptsize},
    ytick={0,1,2,3},
    ylabel={Rel.\ gain in Hit@50 (\%)},
    ylabel style={font=\scriptsize, yshift=-3pt},
    axis x line*=bottom, axis y line*=left,
    ymajorgrids, grid style={evalgrid},
    tick style={black, line width=0.4pt},
    legend style={font=\scriptsize, draw=none, fill=none, at={(0.98,0.14)},
                  anchor=south east, row sep=-2pt, inner sep=1pt},
    legend cell align=left,
    clip=false,
]
\fill[evalshade] (axis cs:5.5,-0.4) rectangle (axis cs:9.4,3.3);
\draw[black!40, line width=0.4pt] (axis cs:-0.4,0) -- (axis cs:9.4,0);
\draw[gray, dashed, line width=0.5pt] (axis cs:5.5,-0.4) -- (axis cs:5.5,3.3);
\node[font=\scriptsize, text=gray] at (axis cs:2.5,3.1) {accumulation};
\node[font=\scriptsize, text=gray] at (axis cs:7.45,3.1) {evaluation};
\addplot[evalblue, line width=1pt, mark=*, mark size=1.3pt]
    coordinates {(0,0.1)(1,0.7)(2,1.0)(3,1.6)(4,1.7)(5,2.2)(6,2.4)(7,2.7)(8,2.6)(9,2.9)};
\addlegendentry{EvoRec}
\addplot[evalorange, line width=0.8pt, mark=square*, mark size=1.2pt]
    coordinates {(0,-0.1)(1,0.6)(2,1.1)(3,1.4)(4,1.7)(5,1.6)(6,1.8)(7,1.7)(8,1.9)(9,1.8)};
\addlegendentry{w/o Tree Evolution}
\addplot[evalgreen, line width=0.8pt, mark=triangle*, mark size=1.4pt]
    coordinates {(0,0.2)(1,0.8)(2,1.1)(3,1.5)(4,1.9)(5,2.1)(6,2.1)(7,1.8)(8,1.8)(9,1.5)};
\addlegendentry{w/o \texttt{ERROR-MEMORY}}
\end{axis}
\end{tikzpicture}
\caption{Daily relative Hit@50 improvement over w/o Reflection on all users. The dashed line separates the accumulation and evaluation periods.}
\Description{Line chart of relative Hit@50 improvement from day D-5 to D+4 for EvoRec, w/o Tree Evolution, and w/o ERROR-MEMORY.}
\label{fig:reflection_rounds}
\end{figure}
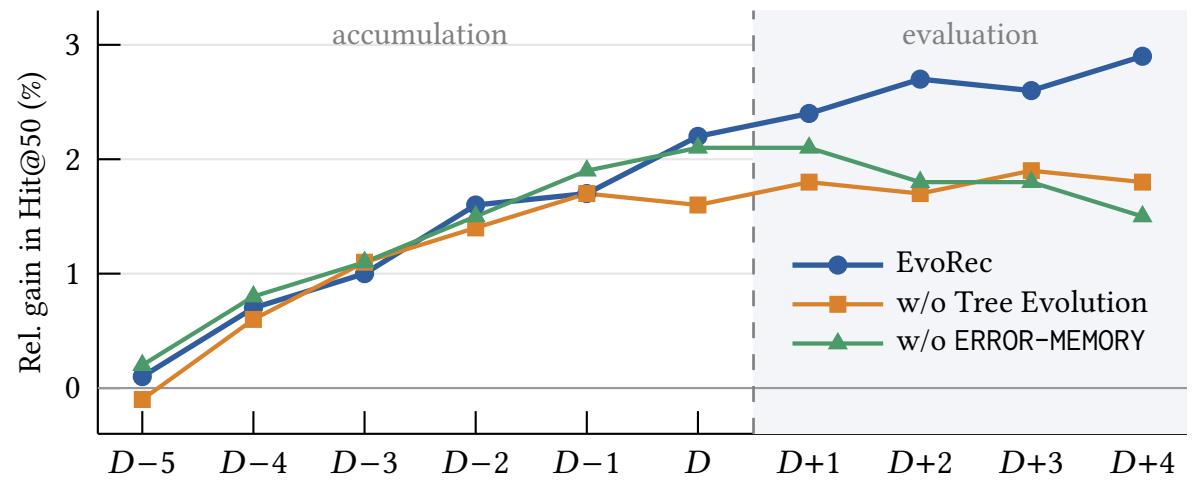

\begin{table}[t]
    \centering
    \caption{Online A/B results within the deployment scope.}
    \label{tab:online_ab}

    \begingroup
    \setlength{\tabcolsep}{3pt}
    \renewcommand{\arraystretch}{1.1}

    \begin{tabular*}{\columnwidth}{
        @{\extracolsep{\fill}}lcc@{}
    }
        \toprule
        \textbf{Metric}
        & \textbf{All users}
        & \textbf{High-intent users} \\
        \midrule

        Impressions
        & +0.8\%
        & \textemdash \\

        Cost
        & +3.6\%
        & \textbf{+5.7\%} \\

        Revenue
        & +4.6\%
        & \textbf{+14.5\%} \\

        CTCVR
        & {+3.5\%}
        & \textbf{+5.6\%} \\

        \bottomrule
    \end{tabular*}
    \endgroup
\end{table}

To examine performance across business categories, we analyze the eight top-level industry categories with the highest revenue,
ordered by revenue and anonymized as C1--C8.
As shown in Figure~\ref{online_by_industry}, C1 achieves a 13.1\% relative improvement in CTCVR, while C3 and C4 improve by 283.1\% and 85.7\%, respectively.
These results show positive effects across multiple core
business categories. EvoRec is integrated into the existing recommendation pipeline
as an independent retrieval channel, without additional training
of the downstream pre-ranking and ranking models.
The online improvements in C1 and other categories under this
deployment setting suggest that candidates retrieved through
explicit user-need reasoning can complement existing ranking
capabilities.

\subsection{Further Analysis}
\label{sec:analysis}

\textbf{Evolution over Rounds.}
Since tree evolution and retraction mainly govern the long-term quality of reflective experience, we track their effects across days in Figure~\ref{fig:reflection_rounds}.
During the first days, gains remain close to zero and occasionally slightly negative, since the tree starts with only its top-level nodes and most repair strategies are not yet verified; the three variants therefore perform similarly.
As experience accumulates, the gain of EvoRec on all users rises to 2.4\% on day $D+1$ and reaches 2.9\% by day $D+4$, as more users accumulate working memory; among users with working memory on day $D+1$, the gain increases from 10.2\% for those with one repair strategy to 18.5\% for those with two or more.
In contrast, the gain of w/o Tree Evolution plateaus at around 1.8\%, and that of w/o \texttt{ERROR-MEMORY} declines from 2.1\% on day $D$ to 1.5\% on day $D+4$ as misleading strategies accumulate, indicating that both mechanisms become more important as reflection proceeds.

\begin{figure}[t]
    \centering
    \includegraphics[width=\linewidth]{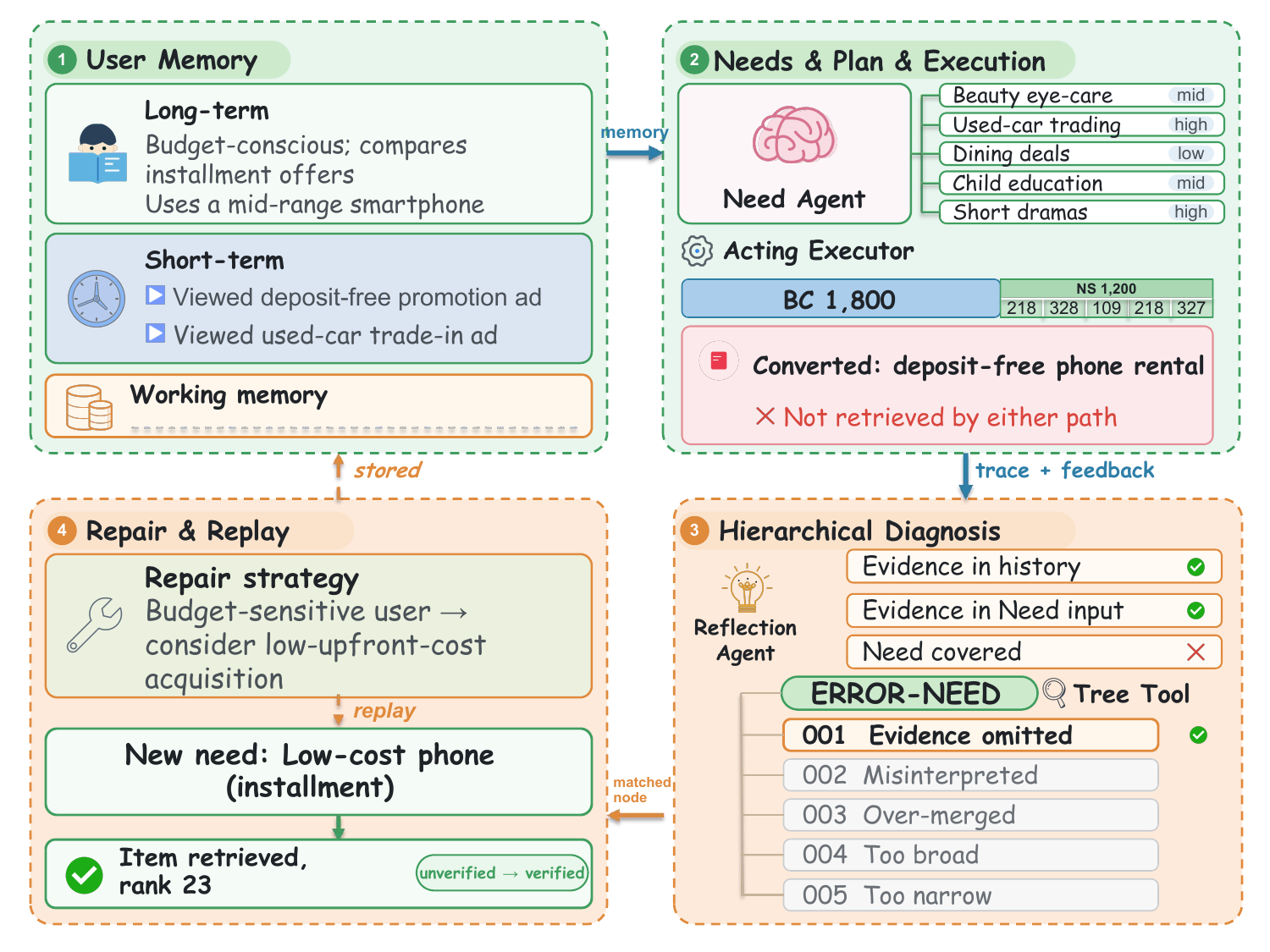}
    \caption{One reflection round (clockwise from top left). BC/NS: Behavior Continuation/Need Satisfaction. User memory is desensitized and abridged.}
    \Description{A four-panel case study showing user memory, inferred needs and execution, hierarchical error diagnosis, and repair with replay.}
    \label{fig:case_study}
\end{figure}

\textbf{Case Study.} Figure~\ref{fig:case_study} shows one reflection round. Although the user's memory indicates budget sensitivity and a recent view of a deposit-free promotion ad, no inferred need reflects this evidence, and neither path retrieves the subsequently converted deposit-free phone rental. Since the evidence appears in the Need Agent's input, the Reflection Agent localizes the error to \texttt{ERROR-NEED} and matches it to \textit{Evidence omitted}.
The repair strategy links budget sensitivity to low-upfront-cost acquisition without naming any item; in replay, the Need Agent combines it with the installment and deposit-free signals in memory to infer a low-cost phone need, the item enters the top-$K$, and the node is marked as verified.
%
%
%
%
%

\section{Conclusion}
\label{sec:conclusion}

We presented EvoRec, a unified agentic recommendation framework with hierarchical reflection.
EvoRec lets agents make recommendation decisions end to end: the Need Agent infers structured user needs from user memory, the Planning Agent allocates the retrieval budget between Behavior Continuation and Need Satisfaction and selects retrieval tools and fusion weights, and the Acting Executor deterministically carries out the plan.
The Dynamic Reflection Tree localizes errors to need inference, planning, or previously applied repair strategies, supplies personalized repair strategies through working memory, and abstracts verified experience into generalized error patterns shared across users, enabling self-evolution without updating agent parameters.
Experiments on Kuaishou show that adaptive decisions outperform static configurations of the same tools, that reflection brings further gains to users with accumulated experience, and that EvoRec improves advertising revenue and conversion in online deployment.
Since reflection currently benefits only users with prior feedback, future work will apply generalized repair strategies to users without working memory and extend EvoRec to other recommendation domains with different tool sets.

%
%
%
%
%

%
%
%
%


\bibliographystyle{ACM-Reference-Format}
\bibliography{references}

\appendix

\clearpage

\section{Implementation Details}
\label{app:implementation}

\begin{table*}[t]
    \centering
    \caption{Effects of retrieval-tool ablation on accuracy
    and candidate diversity.
    Hit@$K$ is reported in percent; other metrics use their original scales. Bold and underlined values indicate
    the best and second-best results, respectively.}
    \label{tab:recall_ablation}

    \begingroup
    \normalsize
    \setlength{\tabcolsep}{2pt}
    \renewcommand{\arraystretch}{1.15}

    \begin{tabular*}{\textwidth}{
        @{\extracolsep{\fill}}l*{12}{c}@{}
    }
        \toprule

        \multirow{2}{*}{\textbf{Configuration}}
        & \multicolumn{6}{c}{\textbf{Hit@$K$} $\uparrow$}
        & \multirow{2}{*}{\textbf{MRR} $\uparrow$}
        & \multicolumn{3}{c}{\textbf{NDCG@$K$} $\uparrow$}
        & \multicolumn{2}{c}{\textbf{Diversity@1000}} \\

        \cmidrule(lr){2-7}
        \cmidrule(lr){9-11}
        \cmidrule(lr){12-13}

        & \textbf{@10} & \textbf{@50} & \textbf{@100}
        & \textbf{@500} & \textbf{@1000} & \textbf{@3000}
        &
        & \textbf{@5} & \textbf{@10} & \textbf{@20}
        & \textbf{Avg. sim.} $\downarrow$
        & \textbf{ILD} $\uparrow$ \\

        \midrule

        \rowcolor{black!6}
        \textbf{All tools}
        & \textbf{8.97} & \textbf{20.02} & \textbf{25.72}
        & \textbf{37.68} & \textbf{41.30} & \textbf{52.55}
        & \textbf{0.0419}
        & \textbf{0.0365} & \textbf{0.0466} & \textbf{0.0570}
        & 0.4801 & 0.5199 \\

        \midrule

        w/o Item-CF
        & 8.32 & 17.59 & 22.05 & 32.15 & 36.29 & 46.94
        & 0.0409
        & \underline{0.0362} & 0.0450 & 0.0539
        & 0.4998 & 0.5002 \\

        w/o User-CF
        & 6.82 & 18.01 & 23.85 & 35.07 & 38.81
        & \underline{51.85}
        & 0.0296
        & 0.0234 & 0.0322 & 0.0421
        & 0.5030 & 0.4970 \\

        \midrule

        w/o Embedding
        & 8.70
        & 19.18 & 24.47 & 35.49
        & 38.62
        & 48.26
        & 0.0410
        & 0.0354 & 0.0452 & 0.0550
        & \textbf{0.4740} & \textbf{0.5260} \\

        w/o BM25
        & \underline{8.84}
        & \underline{19.63} & \underline{25.18} & \underline{36.77}
        & \underline{40.23}
        & 51.09
        & \underline{0.0414}
        & 0.0360 & \underline{0.0459} & \underline{0.0560}
        & \underline{0.4790} & \underline{0.5210} \\

        \midrule

        w/o OneLatent
        & 8.66
        & 19.41 & 24.79 & 36.58
        & 40.18
        & 51.31
        & 0.0405
        & 0.0352 & 0.0450 & 0.0551
        & 0.4835 & 0.5165 \\

        \bottomrule
    \end{tabular*}
    \endgroup
\end{table*}

\subsection{LLM Inference}
All agents are prompted without fine-tuning to produce structured JSON outputs, with thinking mode disabled for the Need and Planning Agents and enabled for the Reflection Agent, temperature 0.2, and vLLM serving on 8 NVIDIA GPUs.
User memory is truncated to the most recent 200 interactions and a long-term profile of at most 1,000 tokens.

\subsection{Retrieval, Ranking, and Reflection}
EvoRec retrieves $B=3{,}000$ candidates per user from five tools: Item-CF, User-CF, embedding-based retrieval, BM25, and OneLatent.
A shared, frozen Qwen3-Embedding-8B~\cite{zhang2025qwen3} encoder produces 4,096-dimensional item representations for both embedding-based retrieval and the Two-Tower ranker; embedding-based retrieval also fills the remaining budget when the selected tools return fewer candidates than their budgets.
SASRec is a Transformer with 2 layers, 2 attention heads, dropout of 0.2, and a maximum sequence length of 100, trained with Adam using a learning rate of $10^{-4}$, a batch size of 512, and early stopping.
Two-Tower takes gated profile embeddings as user inputs and projects both towers into a 64-dimensional space, trained with temperature-scaled sampled softmax ($\tau=0.05$) using one positive conversion item and 50 uniformly sampled negatives with popularity correction per example.
Both rankers are selected by validation NDCG@20.
Ranking fusion follows Eq.~(\ref{eq:acting_rrf}) with $c=60$ and a fixed Base weight of 1.0; the fixed-fusion variants use $\alpha=0.5$ for all users.
For reflection, the attribution cutoff is $K=50$, and at most 5 feedback items are diagnosed per user per day, with conversions prioritized over clicks and recent items prioritized within each type.
A diagnosed error is considered mitigated if the feedback item enters the top-$K$ after replay, in which case the user-level node is marked as verified; an unverified node is promoted once a recurrence of the same error is mitigated under its strategy in a later round.
Abstraction is triggered once 20 verified pending nodes accumulate under the same top-level node, and Merge, Split, and Reparent are applied only after checking that the nodes involved describe a consistent error mechanism and that the resulting hierarchy contains no redundant sibling nodes.

\subsection{Baselines}
\label{app:baselines}
All baselines are trained on data up to day $D$ and evaluated on the same users and candidate pool as EvoRec.
SASRec and UltraGCN use \ph{their official implementations}, with hyperparameters tuned on day $D$.
TIGER and OneRec use beam search with a beam size of \ph{200}.

\subsection{Evaluation Metrics}
\label{app:metrics}
We evaluate recommendation accuracy using Hit@$K$ and NDCG@$K$, and semantic diversity using ILD@$K$.
All metrics are computed over the top-$K$ recommendations,
with cutoffs specified in the tables.
Accuracy metrics are averaged across evaluation users.
ILD@$K$ measures the mean pairwise cosine distance between
item embeddings and is averaged over users with at least
two valid embeddings.
Hit is reported as a percentage.
Let $\mathcal{U}$ denote the evaluation users, $\mathcal{G}_u$ the ground-truth items of user $u$, $\mathcal{L}_u^K$ the top-$K$ list, and $\mathcal{L}_{u,r}$ its item at rank $r$.
All metrics are computed per user and macro-averaged over $\mathcal{U}$:
\begin{align}
\mathrm{Hit}@K &= \frac{1}{|\mathcal{U}|}\sum_{u\in\mathcal{U}}\mathbb{I}\big[\mathcal{L}_u^K\cap\mathcal{G}_u\neq\varnothing\big],\\
\mathrm{NDCG}@K &= \frac{1}{|\mathcal{U}|}\sum_{u\in\mathcal{U}}
\frac{\sum_{r=1}^{K}\mathbb{I}[\mathcal{L}_{u,r}\in\mathcal{G}_u]/\log_2(r+1)}
{\sum_{r=1}^{\min(K,|\mathcal{G}_u|)}1/\log_2(r+1)},\\
\mathrm{MRR}@K &= \frac{1}{|\mathcal{U}|}\sum_{u\in\mathcal{U}}\frac{\mathbb{I}\big[\mathcal{L}_u^K\cap\mathcal{G}_u\neq\varnothing\big]}{r_u^\star},
\end{align}
where $r_u^\star$ is the rank of the first ground-truth item in $\mathcal{L}_u^K$.
ILD@$K$ is one minus the mean pairwise cosine similarity between item embeddings in $\mathcal{L}_u^K$, averaged over users with at least two valid embeddings; Avg.\ Sim.@$K$ equals $1-\mathrm{ILD}@K$.

\begin{table*}[t]
    \centering
    \caption{Comparison of ranking strategies. Base: snake-order merge; +: RRF fusion with Base; TT: Two-Tower. Fixed uses a balanced preset for all users, whereas agentic selects presets per user. Hit@$K$ is reported in percent; other metrics use their original scales. Bold and underlined values indicate the best and second-best results in each column, respectively.}
    \label{tab:ranking_fusion}

    \begingroup
    \setlength{\tabcolsep}{3pt}
    \renewcommand{\arraystretch}{1.15}

    \begin{tabular*}{\textwidth}{
        @{\extracolsep{\fill}}l*{12}{c}@{}
    }
        \toprule
        \multirow{2}{*}{\textbf{Strategy}}
        & \multicolumn{6}{c}{\textbf{Hit@$K$} $\uparrow$}
        & \multirow{2}{*}{\textbf{MRR} $\uparrow$}
        & \multicolumn{3}{c}{\textbf{NDCG@$K$} $\uparrow$}
        & \multicolumn{2}{c}{\textbf{ILD@$K$} $\uparrow$} \\

        \cmidrule(lr){2-7}
        \cmidrule(lr){9-11}
        \cmidrule(lr){12-13}

        & \textbf{@1}
        & \textbf{@5}
        & \textbf{@10}
        & \textbf{@20}
        & \textbf{@50}
        & \textbf{@100}
        &
        & \textbf{@5}
        & \textbf{@10}
        & \textbf{@20}
        & \textbf{@50}
        & \textbf{@100} \\
        \midrule

        Base
        & 1.76 & 5.77 & 8.97
        & 13.24 & 20.02 & 25.72      
        & 0.0419
        & 0.0365 & 0.0465 & 0.0570
        & 0.5373 & 0.5412 \\

        SASRec
        & 1.40 & 4.79 & 7.38
        & 10.86 & 17.13 & 23.29
        & 0.0350
        & 0.0302 & 0.0383 & 0.0468
        & \textbf{0.5564} & \textbf{0.5649} \\

        Two-Tower
        & 1.25 & 4.16 & 6.18
        & 9.02 & 13.82 & 18.65
        & 0.0299
        & 0.0262 & 0.0325 & 0.0394
        & 0.4836 & 0.4852 \\
        \midrule

        + SASRec
        & 1.81 & 6.22 & 9.85
        & 14.41 & 21.51 & 28.19
        & 0.0450
        & 0.0389 & 0.0503 & 0.0616
        & \underline{0.5392} & \underline{0.5528} \\

        + SASRec + TT (fixed)
        & \underline{2.02}
        & \underline{7.26}
        & \underline{11.10}
        & \underline{15.85}
        & \underline{23.28}
        & \underline{29.17}
        & \underline{0.0518}
        & \underline{0.0461}
        & \underline{0.0582}
        & \underline{0.0699}
        & 0.5355
        & 0.5471 \\

        \rowcolor{black!6}
        \textbf{+ SASRec + TT (agentic)}
        & \textbf{2.33}               
        & \textbf{7.52}
        & \textbf{11.37}
        & \textbf{16.23}              
        & \textbf{23.90}
        & \textbf{30.38}              
        & \textbf{0.0525}
        & \textbf{0.0479}
        & \textbf{0.0600}
        & \textbf{0.0720}             
        & 0.5339                      
        & 0.5438 \\                   

        \bottomrule
    \end{tabular*}
    \endgroup
\end{table*}

\section{Additional Experimental Results}

\subsection{Retrieval Tool Collaboration}
Table~\ref{tab:recall_ablation} extends the main-text analysis
across retrieval cutoffs and diversity metrics.
The full configuration achieves the best results on all
reported accuracy metrics.
Removing User-CF produces the largest MRR drop,
from 0.0419 to 0.0296, whereas removing Item-CF causes
the largest Hit@100 drop, from 25.72\% to 22.05\%,
suggesting distinct contributions to early and deeper retrieval.
Removing Embedding reduces Hit@50 by 4.2\% but Hit@3000 by 8.2\%,
from 52.55\% to 48.26\%,
highlighting its contribution at larger retrieval depths.
The tools also differ in their effects on diversity:
removing Item-CF or User-CF lowers ILD@1000 from 0.5199
to 0.5002 and 0.4970, respectively, whereas removing Embedding
raises it to 0.5260.
Thus, the accuracy gains from tool collaboration do not
uniformly translate into greater candidate diversity.
These findings support coordinating complementary retrieval
tools while accounting for their different effects on
retrieval depth and diversity.

\subsection{Ranking Signal Fusion}
Table~\ref{tab:ranking_fusion} extends the main-text analysis
by comparing accuracy and diversity across ranking strategies.
With the same three ranking signals, agentic fusion selects
fusion weights per user rather than applying a shared fixed configuration.
It improves all reported accuracy metrics over fixed fusion,
raising Hit@50 from 23.28\% to 23.90\% and NDCG@10
from 0.0582 to 0.0600.
These gains support adapting signal fusion to individual users.
Meanwhile, ILD@50 and ILD@100 decrease slightly, from 0.5355 to 0.5339
and from 0.5471 to 0.5438, respectively, while SASRec alone
achieves the highest diversity.
Because the candidate sets remain unchanged, these differences
arise from candidate ordering rather than expanded coverage.

\section{Prompts}
\label{app:prompts}

Table~\ref{tab:need_prompt} summarizes the prompt of the Need Agent.

\begin{table*}[t]
\centering
\caption{Prompt of the Need Agent, translated from Chinese and abridged. Braces denote placeholders filled at inference time. The reflection rules are appended only when the user has working memory.}
\label{tab:need_prompt}
\small
\renewcommand{\arraystretch}{1.15}
\begin{tabularx}{\textwidth}{@{}p{2.7cm}X@{}}
\toprule
\textbf{Component} & \textbf{Content} \\
\midrule
Role &
You are the Need Agent of a local-services advertising recommender. Given the long-term profile, recent active behaviors, and current context, output unified needs. Every need must be supported by real input signals; output an empty list when evidence is insufficient. \\
\midrule
\multicolumn{2}{@{}l}{\textit{Source rules}} \\
Recent behavior &
Derive a specific service direction from a single ad in the active behavior sequence, without borrowing unrelated ads. Clicks and conversions indicate signal strength only, not completed transactions. \\
Explicit intent &
Derive only from search queries, query sub-intents, and inquiry messages, preserving their semantic anchors; profile conditions may serve only as constraints. \\
Local-life demand &
Derive from explicit local-life demands in messages or the profile, without requiring a matching ad behavior. \\
Interest preference &
Derive from explicit interests that map directly to local services; profile and consumption signals alone cannot form a need. \\
Cross-domain &
Derive only from non-advertising signals (video, live-streaming, and e-commerce interests or consumption), bridging in one hop to a new deliverable service, e.g., gaming $\rightarrow$ gaming clubs, cooking $\rightarrow$ local dining, automobiles $\rightarrow$ test drives or insurance. \\
\midrule
Generation and merging &
Each need describes one specific, retrievable service direction rather than a broad category. Signals with the same direction are merged with all sources retained; nested or near-duplicate needs are removed, while distinct directions are kept separate. City, price, time, and service form are written as constraints. Output at most 5 needs, preferring clear sources, strong and recent behaviors, and specific directions. \\
\midrule
Output &
Strict JSON with a list of needs, each containing \texttt{need\_description}, \texttt{evidence\_summary}, \texttt{constraints}, \texttt{priority}, and \texttt{source\_types}. Priority is high, medium, or low: explicit actions and strong or repeated recent behaviors rank higher, while cross-domain inferences are usually low. Non-compliant items are deleted before output. \\
\midrule
Input &
Long-term profile: \{long\_term\_memory\}; context: \{location, season, daypart, holiday\}; explicit expressions: \{queries, sub-intents, messages\}; recent behaviors: \{ad ID, timestamp, action, ad content\}. \\
\midrule
Reflection revision &
Additional input: previous needs and reflection memory. Retrieval and conversion facts must not be altered, and diagnostic hypotheses must be re-checked rather than copied. Unaffected needs are kept unchanged; \texttt{refine} and \texttt{merge} modify only the targeted needs, and \texttt{add} creates one new need only when the converted items clearly indicate a specific, independent, retrievable direction, tagged with the source \texttt{reflection\_feedback}. Generic descriptions such as ``shopping experience'' cannot form needs. The complete need list is output. \\
\bottomrule
\end{tabularx}
\end{table*}


%
%
%
%
%
%
%
%
%
%
%

\end{document}